\documentclass[twocolumn,preprintnumbers,amsmath,amssymb,superscriptaddress,prl,showpacs]{revtex4}
\usepackage{graphicx}
\usepackage{dcolumn}
\usepackage{bm}
\usepackage{epsfig}
\usepackage{color,ulem}

\begin{document}

\title{Rheological hysteresis in soft glassy materials}

\author{Thibaut Divoux}
\affiliation{Universit\'e de Lyon, Laboratoire de Physique, \'Ecole Normale Sup\'erieure de
Lyon, CNRS UMR 5672, 46 All\'ee d'Italie, 69364 Lyon cedex 07, France.}
\author{Vincent Grenard}
\affiliation{Universit\'e de Lyon, Laboratoire de Physique, \'Ecole Normale Sup\'erieure de
Lyon, CNRS UMR 5672, 46 All\'ee d'Italie, 69364 Lyon cedex 07, France.}
\author{S\'ebastien Manneville}
\affiliation{Universit\'e de Lyon, Laboratoire de Physique, \'Ecole Normale Sup\'erieure de
Lyon, CNRS UMR 5672, 46 All\'ee d'Italie, 69364 Lyon cedex 07, France.}
\affiliation{Institut Universitaire de France}
\date{\today}

\begin{abstract}
The nonlinear rheology of a soft glassy material is captured by its constitutive relation, shear stress vs shear rate, which is most generally obtained by sweeping up or down the shear rate over a finite temporal window. For a huge amount of complex fluids, the up and down sweeps do not superimpose and define a rheological hysteresis loop. By means of extensive rheometry coupled to time-resolved velocimetry, we unravel the local scenario involved in rheological hysteresis for various types of well-studied soft materials. We introduce two observables that quantify the hysteresis in macroscopic rheology and local velocimetry respectively, as a function of the sweep rate $\delta t^{-1}$. Strikingly, both observables present a robust maximum with $\delta t$, which defines a single material-dependent timescale that grows continuously from vanishingly small values in simple yield stress fluids to large values for strongly time-dependent materials. In line with recent theoretical arguments, these experimental results hint at a universal timescale-based framework for soft glassy materials, where inhomogeneous flows characterized by shear bands and/or pluglike flow play a central role.
\end{abstract}
\maketitle

When submitted to external stress, soft glassy materials such as colloidal gels, clay suspensions, concentrated emulsions, and foams, display a fascinating variety of behaviors because the applied strain may disrupt and rearrange the microstructure over a wide range of spatial and temporal scales leading to heterogeneous flow properties \cite{Ovarlez:2009,Schall:2010}. For more than a decade, flow dynamics have been probed by combining standard rheology, e.g. through the determination of the ``constitutive relation'' between the shear stress $\sigma$ and the shear rate $\dot \gamma$, and local structural or velocity measurements \cite{Moller:2008,Goyon:2008}. While much progress has been made on \textit{steady-state} local flow properties \cite{Nordstrom:2010}, \textit{transient} regimes upon shear start-up have been addressed only recently \cite{Divoux:2010,Moorcroft:2011,Martin:2012,Siebenburger:2012}. Still, in practice, it can be argued that any experimental determination of the flow curve $\sigma(\dot \gamma)$ is effectively transient since it is obtained by sweeping up or down $\dot \gamma$ over a finite temporal window. In other words the measured flow curve coincides with the steady-state relation $\sigma(\dot \gamma)$ \textit{only if} the sweep rate is slow enough compared to any intrinsic timescale of the fluid. Otherwise one expects hysteresis loops in $\sigma(\dot \gamma)$ measurements performed by sweeping up then down the shear rate (or vice versa). This phenomenon, known as ``rheological hysteresis,'' has been commonly observed in a host of complex fluids for about 70 years \cite{Mewis:2009,Green:1943}. However, to date, this ubiquitous signature of the interplay between timescales in complex fluids has not been quantitatively studied by means of local measurements. 

In this Letter, we use time-resolved velocimetry to unveil the local scenario involved in rheological hysteresis in various types of well-studied soft materials. Building upon a systematic experimental protocol, we introduce two observables, $A_\sigma$ and $A_v$, that quantify the amplitude of the hysteresis phenomenon as a function of the sweep rate $\delta t^{-1}$ in macroscopic rheology and local velocity respectively. Both $A_\sigma$ and $A_v$ go through a maximum  with $\delta t$, pointing to the existence of a characteristic timescale $\theta$ for the microstructure dynamics. In thixotropic (laponite) suspensions and (carbon black) gels, $\theta$ is of the order of several hundreds of seconds, while it becomes hardly measurable for simple yield stress fluids such as carbopol and concentrated emulsions. Velocity profiles allow us to understand this evolution by clearly differentiating a succession of homogeneous, shear-banded, and arrested flows depending on the fluid and on the sweep rate, thus providing a local interpretation of rheological hysteresis.
 
\textit{Experimental set-up and protocol.-} Experiments are performed in a polished Plexiglas Couette geometry (typical roughness 15~nm, height 28 mm, rotating inner cylinder of radius 24 mm, fixed outer cylinder of radius 25 mm, gap $e=1$~mm) equipped with a home-made lid to minimize evaporation. Rheological data are recorded with a stress-controlled rheometer (MCR 301, Anton Paar). Two flow curves are successively recorded, first by decreasing the shear rate $\dot \gamma$ from high shear ($\dot \gamma_{\rm max}=10^3$~s$^{-1}$) to low shear ($\dot \gamma_{\rm min}=10^{-3}$~s$^{-1}$) through $N=90$ successive logarithmically-spaced steps of duration $\delta t$ each, and then by immediately increasing $\dot \gamma$ back from $\dot \gamma_{\rm min}$ up to the initial value $\dot \gamma_{\rm max}$ following the same $N$ steps in reverse order. In general the downward and upward flow curves,  $\sigma_{\rm down}(\dot\gamma)$ and $\sigma_{\rm up}(\dot\gamma)$ do not coincide and define a \textit{hysteresis loop}. Simultaneously to the flow curves, the azimuthal velocity $v$ is measured as a function of the radial distance $r$ to the rotor, at about 15~mm from the cell bottom, and with a spatial resolution of 40~$\mu$m  by means of ultrasonic velocimetry \cite{Manneville:2004,remark1}. Velocity data are then averaged over the duration $\delta t$ of each shear-rate step at $\dot \gamma$ and the corresponding velocity profiles $v_{\rm down}(\dot \gamma,r)$ and $v_{\rm up}(\dot \gamma,r)$ are normalized by the rotor velocity $v_0\simeq\dot\gamma e$ to allow for a direct comparison of flow properties at widely different shear rates \cite{remark2}.

We checked that at $\dot \gamma_{\rm max}$ velocity profiles are all linear with no fluctuation so that the flow reaches a steady state within the time interval $\delta t$, even for the fastest sweep rates. Therefore starting from high enough shear rates ensures a well-defined and reproducible initial condition. Our choice of $\dot \gamma_{\rm min}$ and $N$ results from a compromise between good sampling and reasonable total duration. By monitoring the viscoelastic moduli prior to and after each experiment, we checked that evaporation and/or irreversible (chemical) aging of the sample were negligible even for the slowest sweeps ($\delta t=300$~s, $2N\delta t=15$~h). Finally, approximating our step-like protocol by a continuous sweep, the equivalent sweep rate is ${\rm d}(\log\dot\gamma)/{\rm d}t=1/n\delta t$, where $n=N/\log(\dot \gamma_{\rm max}/\dot \gamma_{\rm min})$ is the number of steps per decade. In the present protocol we fix $n=15$ and identify the sweep rate with $\delta t^{-1}$, while keeping in mind that $n\delta t$ is the actual control parameter (see Fig.~1 in supplemental material).

\begin{figure}[!t]
\begin{center}
\includegraphics[width=7.5cm]{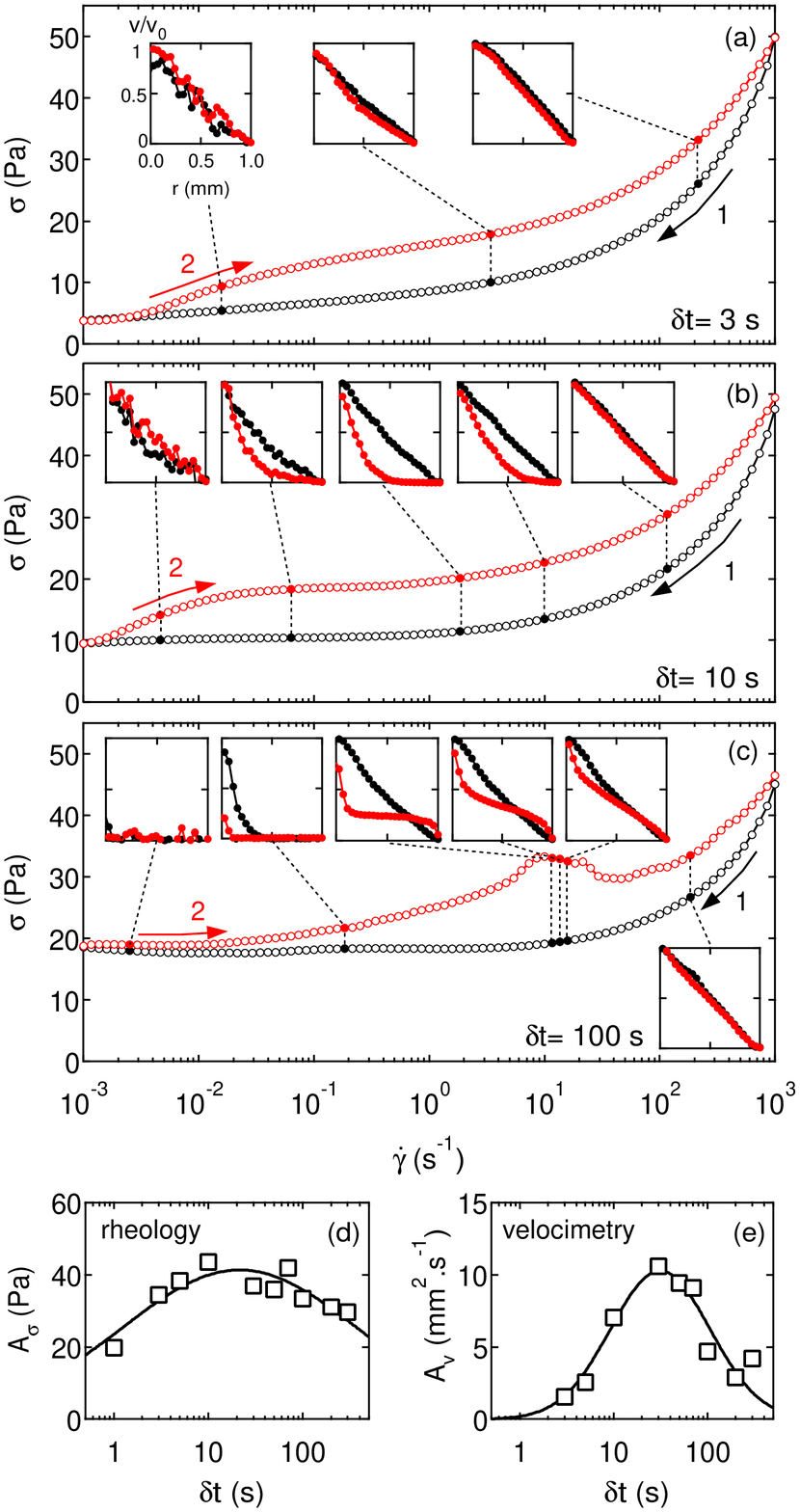}
\end{center}
\caption{\small{(color online). (a)--(c) Flow curves $\sigma$ vs $\dot \gamma$ of a 3\%~wt. laponite suspension obtained by first decreasing $\dot \gamma$ from $10^3$ to $10^{-3}$~s$^{-1}$ in 90 logarithmically-spaced steps of duration $\delta t$ each (black symbols), and then increasing $\dot \gamma$ over the same range (red symbols). Each plot corresponds to a different time interval per step: (a) $\delta t=3$~s, (b) 10~s and (c) 100~s. Insets: velocity profiles inside the gap recorded at the same shear rate during the downward (black) and upward (red) sweeps. Velocity data are normalized by the rotor velocity $v_0$ and the vertical scale goes from 0 to 1. (d)  Hysteresis loop area $A_{\sigma}$ defined by Eq.~(\ref{eq.asigma}) vs $\delta t$. (e) Area $A_v$ defined from the velocity profiles by Eq.~(\ref{eq.av}) vs $\delta t$. Solid lines are lognormal fits of the $A_{\sigma}$ and $A_v$ data.}
}
\label{fig.1}
\end{figure}

\textit{Laponite suspension.-} Clay suspensions are well known to exhibit large rheological hysteresis \cite{Moller:2009b}. Here we focus on laponite samples prepared by mixing ultrapure water with 3\%~wt. of laponite powder (Rockwood, grade RD) and 0.3\%~wt. of hollow glass spheres (Sphericel, Potters) acting as acoustic contrast agents \cite{Manneville:2004}. Figure~\ref{fig.1} shows the hysteresis loops for $\delta t=3$, 10, and 100~s. To quantify their areas, we introduce the following observable:
\begin{equation}
A_{\sigma} \equiv \int_{\dot \gamma_{\rm min}}^{\dot \gamma_{\rm max}}|\Delta \sigma (\dot \gamma)| \,{\rm d}(\log\dot \gamma)\,,
\label{eq.asigma}
\end{equation}
where $\Delta \sigma (\dot \gamma)=\sigma_{\rm up}(\dot\gamma)-\sigma_{\rm down}(\dot\gamma)$. Note that the logarithmic sampling of $\dot\gamma$ gives an equal weight to low and high shear rates. Strikingly, when the sweep rate is decreased, i.e. when $\delta t$ is increased, $A_{\sigma}$ goes through a maximum at $\delta t^\star \simeq 25$~s [Fig.~\ref{fig.1}(d)]. To uncover the local scenario underlying this non-monotonic behavior of $A_{\sigma}$, we turn to time-resolved velocity profiles. In a way similar to Eq.~(\ref{eq.asigma}), we consider $\Delta  v(\dot \gamma,r)= v_{\rm up}(\dot \gamma,r)-v_{\rm down}(\dot \gamma,r)$, and integrate it over the gap and then over the range of explored shear rates:
 \begin{equation}
A_v \equiv \int_{\dot \gamma_{\rm min}}^{\dot \gamma_{\rm max}} \int_{0}^e |\Delta v(\dot \gamma,r)|\, {\rm d}r \,\,{\rm d}(\log\dot \gamma)\,.
\label{eq.av}
\end{equation}
As shown in Fig.~\ref{fig.1}(e), the evolution of this local observable $A_v$ with $\delta t$ directly reflects that of the area $A_\sigma$ extracted from the sole global rheology and its maximum is reached for a similar value of $\delta t^\star \simeq 25$~s. This suggests that the behavior of $A_{\sigma}$ results from the bulk flow properties at the mesoscopic scale. More precisely, for small values of $\delta t$, the laponite suspension is quickly ``quenched'' from high shear rates to lower ones. Velocity profiles  (see also supplementary movies) are all linear both on the way down and on the way up even at the smallest shear rates [Fig.~\ref{fig.1}(a)]: the laponite suspension is not given enough time to restructure and remains fluid throughout the whole cycle, which leads to vanishingly small values of $A_v$.  For intermediate values of $\delta t$, the velocity profiles along the two flow curves now strongly differ [Fig.~\ref{fig.1}(b)]: while they remain linear during the downward sweep, they become inhomogeneous during the upward sweep and exhibit shear banding over a large range of shear rates. Here, the suspension is given enough time to restructure leading to an arrested band subject to physical aging close to the fixed wall at $r=e$, that progressively disappears as the shear rate is increased. This contributes to increase $A_v$. Finally, for large values of $\delta t$, flow arrest also occurs along the downward sweep, which tends to decrease $A_v$ [Fig.~\ref{fig.1}(c)]: shear banding is observed for $\dot\gamma\sim 0.1$--1~s$^{-1}$ until the system experiences total slippage at the rotor (i.e. $v=0$ everywhere in the bulk) for $\dot \gamma \lesssim 10^{-2}$~s$^{-1}$. This fully arrested state persists on the upward sweep up to much higher shear rates ($\dot\gamma\sim 10$~s$^{-1}$) and gives way to a homogeneously sheared state above a small interval of shear rates that corresponds to decreasing shear stresses and to inhomogeneous flows. This scenario is robust and does not significantly depend on the boundary conditions or on sample age since preparation (see Figs.~2 and 3 in supplemental material). As discussed below, we suggest that the characteristic time $\theta=n\delta t^\star\simeq 375$~s results from the competition between physical aging (restructuration) and shear rejuvenation (structure break up) \cite{Coussot:2002}.

\begin{figure}[!t]
\begin{center}
\includegraphics[width=7.5cm]{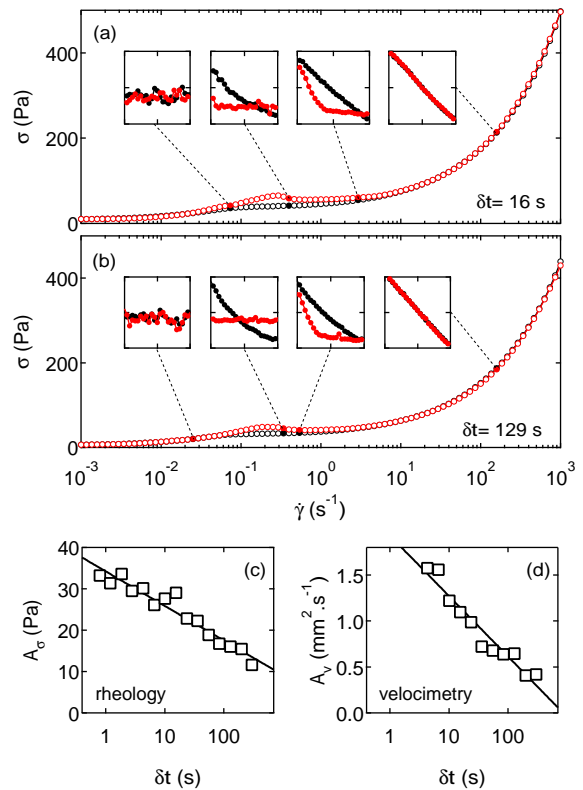}
\end{center}
\caption{\small{(color online). Same as Fig.~\ref{fig.1} for a 1\%~wt. carbopol microgel. Solid lines in (c) and (d) are linear fits of the $A_{\sigma}$ and $A_v$ data in semilogarithmic scales.}
}
\label{fig.2}
\end{figure}

\textit{Carbopol microgel.-}  Let us turn to a simple yield stress fluid, namely a carbopol microgel, where restructuration is expected to be fast \cite{Coussot:2010,Divoux:2012}. Results are reported in Fig.~\ref{fig.2}. The global observable $A_\sigma$ is of the same order as for the laponite suspension \cite{remark3}. Still, in the case of carbopol, $A_\sigma$ is monotonically decreasing over the range of explored $\delta t$ [Fig.~\ref{fig.2}(c)]. Here again the local observable $A_v$ follows the exact same trend [Fig.~\ref{fig.2}(d)]. Velocity profiles reveal that the local scenario remains qualitatively the same whatever the sweep rate: homogeneous shear flow is observed along most of the downward sweep together with an ever-increasing amount of wall slip. For $\dot\gamma\lesssim 0.1$~s$^{-1}$ the local shear rate vanishes and the flow becomes pluglike at roughly half the rotor velocity. On the way up, pluglike flow gives way to shear-banded profiles right after the stress maximum and eventually to linear profiles at high shear rates [see insets in Fig.~\ref{fig.2}(a,b) and movies in the supplemental material], consistently with previous reports on startup experiments at constant shear rate \cite{Divoux:2010,Divoux:2012}. However, as $\delta t$ is increased, the range of shear rates over which shear banding is observed gets narrower, leading to smaller values of $A_v$.

\begin{figure*}[!t]
\begin{center}
\includegraphics[angle=-90,width=17.5cm]{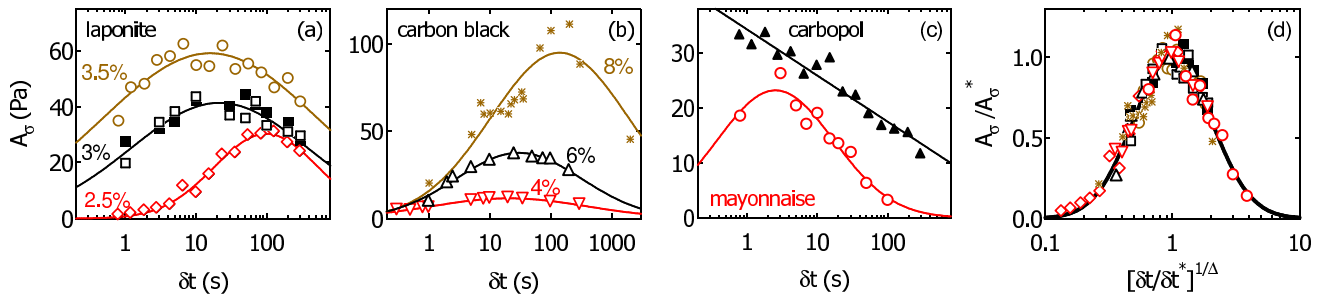}
\end{center}
\caption{\small{(color online). Area $A_\sigma$ of the hysteresis loop as a function of the waiting time per point $\delta t$ for (a) laponite suspensions at 2.5\%~wt. ($\lozenge$), 3\%~wt. ($\Box$), and 3.5\%~wt. ($\circ$), (b) carbon black gels at 4\%~wt. ($\lozenge$), 6\%~wt. ($\Box$), and 8\%~wt. ($\circ$), and (c) a carbopol microgel at 1\%~wt ($\Box$) and a commercial mayonnaise (Casino) ($\circ$). The $\blacksquare$ symbols in (a) correspond to a 3\%~wt. laponite suspension free of acoustic contrast agents and show that the addition of hollow glass spheres does not affect $A_\sigma$. In (b), carbon black particles and 1\%~wt. hollow glass spheres are suspended in a light mineral oil (Sigma, density 0.838, viscosity 20~mPa\,s), which allows experiments over 100~h ($\delta t=2\cdot10^3$~s) without any evaporation. (d) Normalized $A_\sigma/A_\sigma^*$ data for all fluids except carbopol vs $[\delta t/\delta t^*]^{1/\Delta}$, where $A_\sigma^\star$, $\delta t^\star$, and $\Delta$ are the parameters of the lognormal fits, $A_\sigma=A_\sigma^\star\exp\{-[\log(\delta t/\delta t^\star)/\Delta]^2\}$, displayed as solid lines in (a)--(c).}
}
\label{fig.3}
\end{figure*}

\textit{Discussion.-} To further establish the generality of our results we explore a broader range of soft glassy materials in Fig.~\ref{fig.3}. For three different concentrations, thixotropic laponite suspensions [Fig.~\ref{fig.3}(a)] and carbon black gels [Fig.~\ref{fig.3}(b)] display a bell-shaped $A_\sigma$. The case of a concentrated emulsion [commercial mayonnaise, Fig.~\ref{fig.3}(c)] seems intermediate between carbopol and the two previous materials: the loop area $A_\sigma$ exhibits a maximum at the faster end of the accessible range of sweep rates corresponding to $\delta t^\star\simeq 2.5$~s i.e. $\theta\simeq 40$~s. These bell-shaped curves are well captured by lognormal fits, leading to the master curve shown in Fig~\ref{fig.3}(d). 

The above results hint at a unified picture in which the behaviors of both $A_\sigma$ and $A_v$ are interpreted in terms of a single characteristic timescale $\theta=n \delta t^\star$. For $n\delta t<\theta$, the bulk material remains fluidized during the downward sweep while flow heterogeneity during the upward sweep keeps increasing with $n\delta t$ [Fig.~\ref{fig.1}(a,b)], leading to increasing $A_v$ and $A_\sigma$. Note that $A_v$ displays a sharper increase than $A_\sigma$. The lower sensitivity of $A_\sigma$ may be attributed to microstructural changes in the {\it fluidized} suspension (e.g. local fluctuations in the concentration or in the size of colloidal aggregates) responsible for differences in viscosity only, so that downward and upward flow curves may not superimpose ($A_\sigma\neq 0$) although velocity profiles are all linear ($A_v\simeq 0$). Making a rigorous connection between $A_\sigma$ and $A_v$ from theoretical approaches and/or material-dependent modeling is a challenging task that must be addressed in the future together with time-resolved structural characterization during the sweeps. 

As for the maximum of $A_\sigma$ and $A_v$, velocity data show that $\theta$ corresponds to the point where flow heterogeneity first appears during the downward sweep. $\theta^{-1}$ can thus be interpreted as the sweep rate for which restructuration balances shear-induced fluidization over the cycle. For $n\delta t>\theta$ the decrease of $A_\sigma$ is easily explained if one notices that the shear stress becomes independent of the sweep direction as soon as fully arrested states are reached [Figs.~\ref{fig.1}(c) and \ref{fig.2}(a,b)]. Whatever the material, arrested states span over larger portions of the flow curve as $n\delta t$ is increased at the expense of transient shear-banded flows. This tends to decrease the low shear rate contribution to the loop area, hence the decreasing behavior of $A_\sigma$ and $A_v$. In other words, the slower the sweep rate, the closer the flow gets to its steady state, and thus the smaller the areas $A_\sigma$ and $A_v$ as expected from first intuition \cite{Mewis:2009}.

We emphasize that wall slip, i.e. the presence of lubricating layers at the boundaries, does not seem to bring any significant contribution to $A_\sigma$ and $A_v$. Indeed using roughened walls does not affect the shape of the areas nor the value of $\theta$ (see supplementary figure~2). Total slippage thus appears as a mere consequence of flow arrest in the bulk and the nonmonotonic behavior of $A_\sigma$ and $A_v$ originates only from the bulk flow heterogeneity. Still potential effects of the flow geometry on the shape of the hysteresis loops and/or on $\theta$ should  be investigated. Interestingly, in both carbopol microgels and laponite suspensions, inhomogeneous shear-banded flows are observed concomitantly with sections of the upward flow curves where $\sigma$ presents a plateau [Fig.~\ref{fig.1}(b)] or decreases with increasing $\dot\gamma$ [Fig.~\ref{fig.1}(c) and \ref{fig.2}(a,b)]. This is in agreement with a universal criterion proposed recently in Ref.~\cite{Moorcroft:2012} which predicts transient shear banding to arise just after any stress overshoot during startup.

Finally Fig.~\ref{fig.3} suggests that the characteristic time $\theta$ continuously grows when going from simple yield stress fluids (carbopol, emulsions) to highly time-dependent materials (laponite, carbon black). In carbopol microgels $\theta$ is too short to be measured and only the decreasing parts of $A_\sigma$ and $A_v$ are observed. For thixotropic materials $\theta$ could be linked to the history-dependent timescales for structure breakdown and build-up inferred from shear rate jumps \cite{Dullaert:2005}. However, here, $\theta$ is protocol-independent as long as $\dot \gamma_{\rm min}$ and $\dot \gamma_{\rm max}$ are well separated. We may therefore speculate that it relates to the intrinsic material ``restructuring time'' invoked in recent theoretical arguments and numerical simulations \cite{Coussot:2010,Martens:2012}, e.g., the (very short) duration of rearrangement events in simple yield stress fluids (known as T1 events in emulsions and foams \cite{Biance:2009}) and the (much longer) aggregation time of colloidal particles in other systems \cite{Mongondry:2005}. The observed decrease of $\theta$ with laponite concentration would thus result from faster aggregation dynamics while the inverse trend in carbon black gels could be attributed to sedimentation of large clusters which gets slower as the concentration is increased.

To conclude our study provides experimentalists with observables that quantify the distance to steady state in measurements of the constitutive relation: whatever the soft glassy system, sweep rates much larger than $\theta^{-1}$ should be used to ensure that effects of long-lived transients and inhomogeneous shear-banded or arrested flows are minimized. Exploring the hysteresis maximum through faster sweeps provides interesting physical insights on how local flow properties drive rheological hysteresis and on the type of material, with an apparently continuous transition from strong thixotropy (large $\theta$) to simple yield stress fluids (vanishingly small $\theta$). Clearly, spatially-resolved models are needed to further explain and complete our experimental findings. Soft-glassy rheology approaches \cite{Moorcroft:2011,Moorcroft:2012} and local fluidity models \cite{Martens:2012,Rogers:2008} appear as ideal candidates to check whether rheological hysteresis could be envisioned in a universal framework as suggested by the present work.

\begin{acknowledgments}
We thank Y. Forterre and V. Trappe for providing the carbopol and the carbon black, and C.~Barentin and G.~Ovarlez for fruitful discussions. SM acknowledges funding from the European Research Council under the European Union's Seventh Framework Programme (FP7/2007-2013) / ERC grant agreement n$^\circ$~258803. 
\end{acknowledgments}

\clearpage
\newpage
\setcounter{figure}{0}

\section{Supplemental material}
\begin{center}
{\bf Rheological hysteresis\\ in soft glassy materials}
\end{center}

\subsection{Supplementary movies} 

Five movies are provided as supplemental material to show the full data sets corresponding to Figs. 1 and 2 in the main text. The files names indicate the system under investigation (laponite suspension or carbopol microgel) and the value of the waiting time per point $\delta t$. Velocity profiles are shown only for $\dot\gamma < 250$~s$^{-1}$ due to the limitation in the ultrasonic pulse repetition frequency to 20~kHz, which sets an upper bound on the measurable velocities \cite{Manneville:2004}. At very small shear rates noise becomes too large for velocities to remain reliable. Indeed, the amount of velocity data recorded at each shear-rate step of fixed duration $\delta t$ increases with $\dot \gamma$ \cite{Manneville:2004}. Therefore, at small $\dot \gamma$, the lower statistics result in significant noise in the velocity profiles.

In the movies, the data corresponding to the downward (upward resp.) sweep are shown in red (blue resp.). The velocities profiles measured during the downward sweep are replotted with smaller symbols together with the upward sweep data, in order to allow for a direct comparison of the two velocity profiles obtained at the same shear rate. Black squares at $r=0$ and $r=1$~mm indicate respectively the rotor velocity ($v_0\simeq \dot\gamma e$) and the stator velocity ($v=0$).

\subsection{Influence of the number of steps within a cycle} 

\begin{figure}[h!]
\begin{center}
\includegraphics[width=8cm]{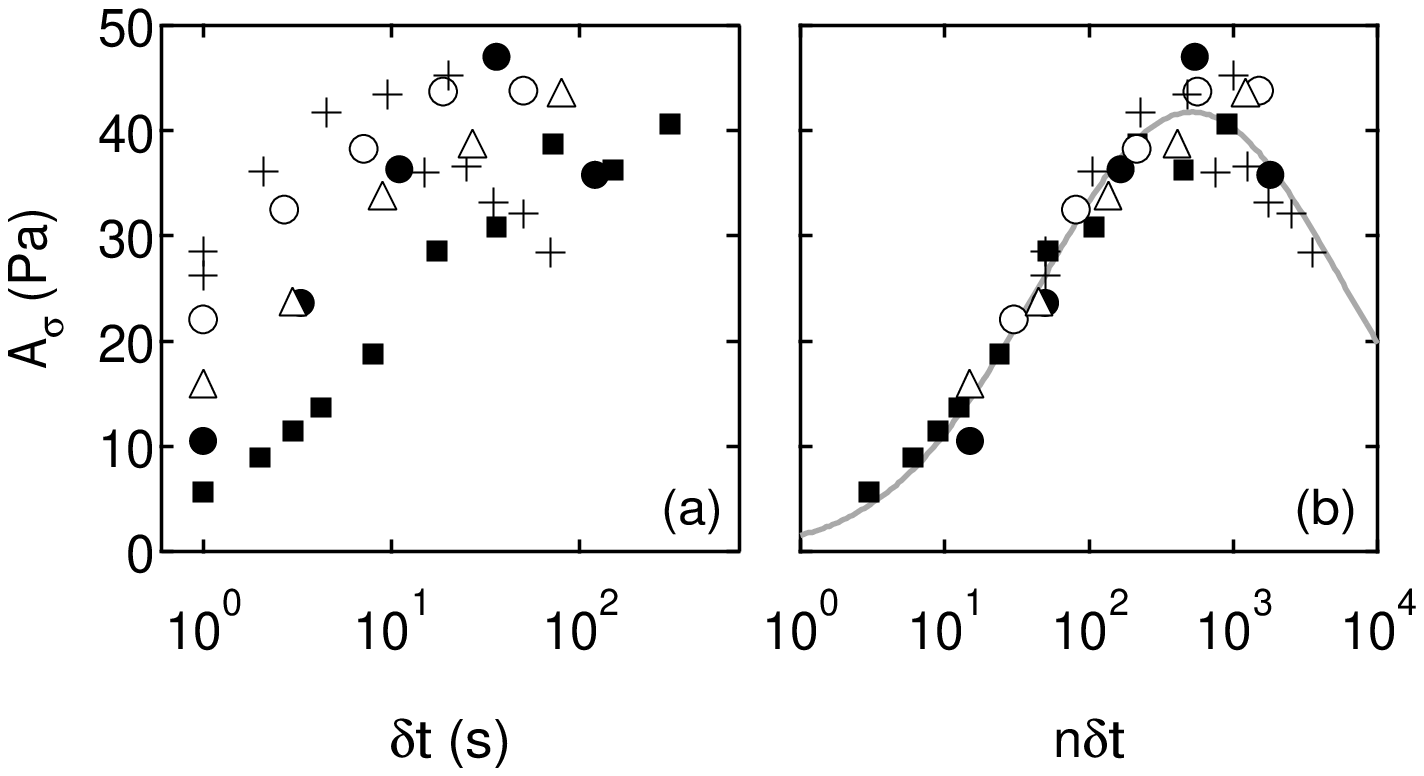}
\end{center}
\caption{\small{(a) Area $A_\sigma$ of the hysteresis loop as a function of the waiting time per point $\delta t$ for different numbers of measurement points per decade $n=3$ ($\blacksquare$), $8$ ($\bullet$), $15$ ($\vartriangle$), $30$ ($\circ$), and $50$ ($+$). (b) Same data plotted as a function of $n\delta t$. The gray curve is a lognormal fit of the rescaled data. Data obtained on a 3.5\%~wt. laponite suspension in a smooth Couette cell. }
}
\label{suppfig.1}
\end{figure}

Figure~\ref{suppfig.1} shows the curves $A_\sigma$ vs $\delta t$ obtained on a laponite suspension for different values of $n$, the number of measurement points per decade in the shear rate sweeps. The data all nicely collapse on a single curve when plotted as a function of $n\delta t$. In the experiments reported in the main text $n$ is fixed arbitrarily to 15 and only $\delta t$ is varied. The collapse shown in Fig.~\ref{suppfig.1}(b) indicates that the true control parameter is $n\delta t$. Therefore, when interpreting the time interval $\delta t^\star$ at which $A_\sigma$ reaches a maximum, one has to keep in mind that the corresponding physical timescale is rather $\theta=n\delta t^\star\simeq 500$~s in the case of Fig.~\ref{suppfig.1}.

\subsection{Influence of the boundary conditions}

\begin{figure}[h!]
\begin{center}
\includegraphics[width=8cm]{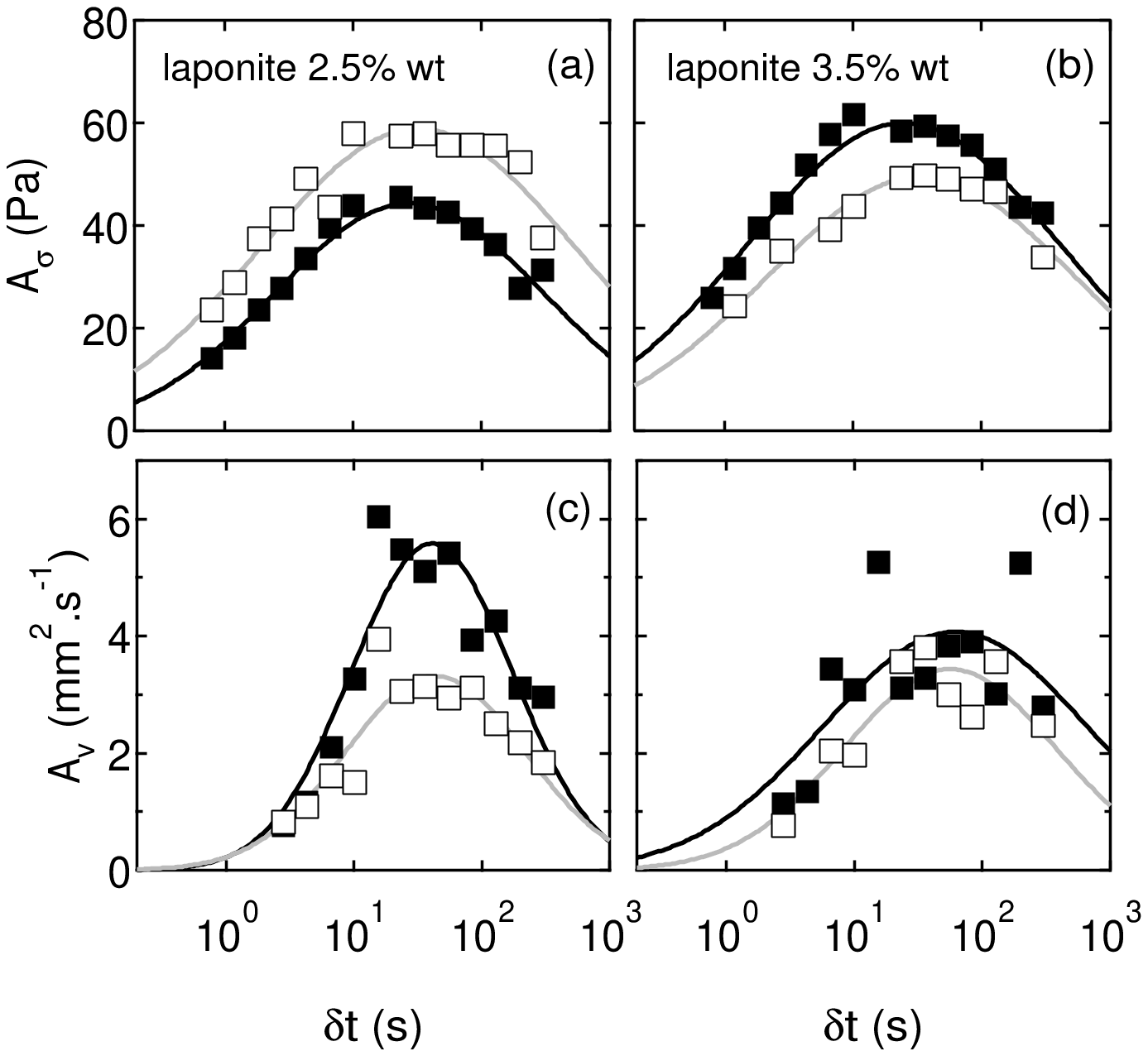}
\end{center}
\caption{\small{(a,b) Area $A_\sigma$ of the hysteresis loop as a function of the waiting time per point $\delta t$ in smooth ($\Box$) and rough ($\blacksquare$) boundary conditions and (c,d)~corresponding area $A_v$ deduced from the velocity profiles for laponite suspensions at (a,c)~2.5\%~wt. and (b,d)~3.5\%~wt. Solid lines are lognormal fits.}
}
\label{suppfig.2}
\end{figure}

We checked for the influence of the boundary conditions on both observables $A_\sigma$ and $A_v$. Figure~\ref{suppfig.2} reports results obtained on two different laponite suspensions with smooth and rough boundary conditions. ``Smooth'' walls refer to polished Plexiglas, as used throughout the main text. The typical roughness of smooth walls is 15~nm. ``Rough'' boundary conditions refer to a sandblasted Plexiglas Couette cell  (roughness of about 1~$\mu$m) with the same dimensions as the smooth cell. We could not detect any systematic influence of the boundary conditions on $A_\sigma$ and $A_v$. In particular, the positions of the maxima $\delta t^\star$ in Fig.~\ref{suppfig.2}(a) and (b) are not affected by the wall roughness although the values of the hysteresis area $A_\sigma$ can be up to 30\% larger or smaller over the whole range of investigated $\dot\gamma$ depending on the boundary conditions.

In Fig.~\ref{suppfig.2}(c), $A_v$ is seen to be about twice as large in the rough cell as in the smooth cell. A close inspection of the velocity profiles indicates that the material tends to stick at the stator in a rough cell so that fully arrested profiles (with $v=0$ everywhere) persist for longer times in the upward shear rate sweeps, leading to larger values of $A_v$. The larger scatter in the $A_v$ data of Fig.~\ref{suppfig.2}(d) for rough boundary conditions also seems to arise from more erratic slip phenomena than in smooth conditions. Still the general evolution of the velocity profiles during shear-rate sweeps with rough walls remains otherwise very close to that observed with smooth walls.

\subsection{Influence of the natural aging process}

\begin{figure}[tb]
\begin{center}
\includegraphics[width=6.5cm]{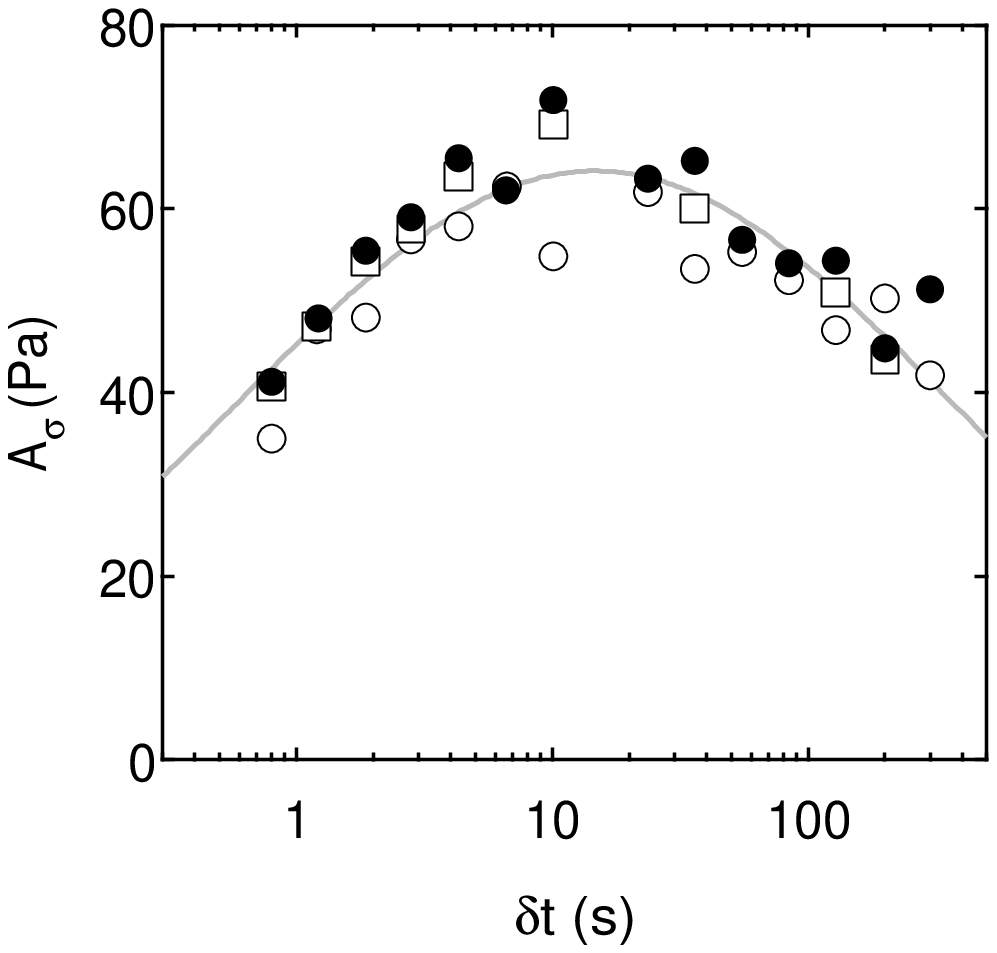}
\end{center}
\caption{\small{Area $A_\sigma$ of the hysteresis loop as a function of the waiting time per point $\delta t$ in a 3.5\%~wt. laponite suspension for different ages $t_w$ since sample preparation: $t_w=1$~day ($\circ$), $10$~days ($\Box$), and $40$~days ($\bullet$). Experiments performed in a smooth Couette cell. The sample is free of acoustic contrast agents. The gray curve is a lognormal fit of the whole data set.}
}
\label{suppfig.3}
\end{figure}

Laponite suspensions are known to exhibit strong aging properties due to slow chemical and physical processes \cite{Ruzicka:2011}. We tested the influence of the sample age on rheological hysteresis by preparing a laponite suspension that was left at rest for a time $t_w$ before part of the sample was loaded into the smooth Couette cell and submitted to series of shear rate sweeps as described in the main text. Figure~\ref{suppfig.3}
confirms that aging processes over more than one month do not significantly affect the quantitative behaviour of $A_\sigma$. 

\begin{thebibliography}{}
\bibitem{remark3} The apparently smaller hysteresis in carbopol (Fig.~\ref{fig.2}) than in laponite (Fig.~\ref{fig.1}) is due to the much larger vertical scale in the flow curves.
\end{thebibliography}
\end{document}